\documentclass{PoS}
\usepackage{textcomp}
\usepackage{amsmath,graphics,graphicx}
\usepackage{latexsym}
\usepackage[authoryear,square]{natbib}
\bibpunct{(}{)}{;}{a}{}{,}

\title{Solar and Heliospheric Physics with the Square Kilometre Array}

\ShortTitle{Solar and Heliospheric Physics with SKA}

\author{
Valery~M.~Nakariakov$^1$,
{Mario M. Bisi}$^2$,
Philippa~K.~Browning$^3$,
Dalmiro Maia$^4$,
Eduard~P.~Kontar$^{5}$,
Divya Oberoi$^{6}$,
Peter~T.~Gallagher$^7$,
Iver~H.~ Cairns$^{8}$,
Heather Ratcliffe$^{1}$
\\
$^1$ Centre for Fusion, Space and Astrophysics, Physics Department, University of Warwick, Coventry CV4 7AL, UK;
$^2$RAL Space, Science \& Technology Facilities Council -- Rutherford Appleton Laboratory, Harwell Oxford, Oxfordshire, OX11 0QX, England, UK;
$^3$ Jodrell Bank Centre for Astrophysics, University of Manchester, Manchester, M13 9PL, UK;
$^4$CICGE, Observatorio Astronomico Professor Manuel de Barros, Faculdade de Ciencias da Universidade do Porto, Vila Nova de Gaia, Portugal;
$^5$ School of Physics and Astronomy, University of Glasgow, Glasgow, G12 8QQ, UK;
$^6$National Centre for Radio Astrophysics, Tata Institute of Fundamental Research, India;
$^7$ School of Physics, Trinity College Dublin, 2, Dublin, Ireland;
$^8$School of Physics, University of Sydney, Sydney, NSW 2006, Australia.
\\

\\
E-mail: \email{V.Nakariakov at warwick.ac.uk; Mario.Bisi at stfc.ac.uk}
}

\abstract{The fields of solar radiophysics and solar system radio physics, or radio heliophysics, will benefit immensely from an instrument with the capabilities projected for SKA.  Potential applications include interplanetary scintillation (IPS), radio-burst tracking, and solar spectral radio imaging with a superior sensitivity.  These will provide breakthrough new insights and results in topics of fundamental importance, such as the physics of impulsive energy releases, magnetohydrodynamic oscillations and turbulence, the dynamics of post-eruptive processes, energetic particle acceleration, the structure of the solar wind and the development and evolution of solar wind transients at distances up to and beyond the orbit of the Earth. The combination of the high spectral, time and spatial resolution and the unprecedented sensitivity of the SKA will radically advance our understanding of basic physical processes operating in solar and heliospheric plasmas and provide a solid foundation for the forecasting of space weather events.}

\FullConference{
Advancing Astrophysics with the Square Kilometre Array\\
June 8-13, 2014\\
Giardini Naxos, Sicily, Italy}

\newcommand{\skipthis}[1]{}
\begin{document}

\section{Introduction}

The Sun is the brightest radio object in the Universe visible from the Earth. In powerful flares, the radio flux density may exceed 10$^9$~Jy. The wide variety of mechanisms, both coherent and incoherent, for solar and heliospheric radio emission provide us with unique information required for understanding the basic physical processes operating in natural and laboratory plasmas, at both microscopic and macroscopic levels. The topics of ongoing intensive investigations are the fundamental problems of plasma astrophysics: the release of magnetic energy, acceleration of charged particles, magnetohydrodynamic (MHD) waves and turbulence, wave-particle interaction, etc. The proximity of the Sun to the Earth allows for its study with an unprecedented combination of time, spatial and spectral resolution, and a unique opportunity to study fundamental plasma physics processes both in situ and remotely.
Finally, plasma physics processes in the solar atmosphere are directly relevant to geophysical challenges such as climate change and space weather; a strong additional motivation for the intensive development of solar and heliospheric radio physics.

Observations of solar and heliospheric radio emission are mainly performed with dedicated instruments, such as radio interferometers. However these
are rather limited. For example the highest spatial resolution in the microwave band currently achieved by the Nobeyama Radioheliograph \citep[NoRH,][]{1994IEEEP..82..705N}, 5" at 34~GHz, is much lower than the spatial scale of plasma structures in the solar corona as resolved in the EUV and X-ray bands (smaller than 1"). Not even the upcoming new generation of state-of-the-art specialised solar radio interferometers, namely the Chinese Spectral Radioheliograph \citep[CSRH, frequency range 0.4--15 GHz, longest baseline 3 km,][]{2009EM&P..104...97Y}, the upgraded Siberian Solar Radio Telescope \citep[SSRT, frequency range 4--8 GHz, longest baseline 622.3~m,][]{2014RAA....14..864L} and the Expanded Owens Valley Solar Array
\citep[e-OVSA, frequency range 1--18 GHz, longest baseline 1.8~km,][]{2012AAS...22020430G}
will reach the SKA's spatial resolution and sensitivity. In short, as well as providing simultaneously high spectral and spatial resolution unavailable with current instruments, the SKA will radically (by two orders of magnitude) improve on their sensitivity, allowing for the detection of a number of physical phenomena predicted theoretically. {The breakthrough potential of SKA in solar and heliospheric studies in the low frequency band has already been demonstrated in frames of the LOw Frequency ARray (LOFAR) and Murchison Widefield Array (MWA), both of
which are SKA pathfinder projects.  These instruments include solar and heliospheric physics, and space weather amongst their key science objectives and have already lead to several interesting publications \citep[e.g.][]{2011pre7.conf..507M, 2011ApJ...728L..27O, 2013PASA...30...31B}.}

A further interesting opportunity is connected with the fact that for a 300~km baseline, the proximity of the Sun to the interferometer puts it in the near-field zone of the instrument at higher frequencies.
The sphericity of the waves coming from spatially localised solar sources can be measured and the radial distance to the source can be estimated, providing us with 3D information: both angular coordinates on the plane-of-the-sky and the distance to the source \citep[e.g. giving radial resolution of 0.1~$R_\odot$ at 1.5~GHz on a 300~km baseline,][]{1997LNP...483..207B}.

For imaging purposes, solar observations are particularly challenging.  First of all there is the immense dynamic range.  During major outbursts the flux can be dominated by very spatially-localised sources
and simultaneously there are elongated features whose brightness temperature over the same spatial extent as the narrow source could be nine orders of magnitude lower.  The spatial scales of those emissions vary widely, with the thermal emission from the corona exceeding the size of the solar disk, and loop gyrosynchrotron emissions reaching even larger sizes, while the emission from noise storms is close to the minimum size defined by scattering in the solar atmosphere {(e.g. the effect of plasma turbulence in the meter wave range, \citep{1998ARA&A..36..131B}) }.  The temporal scales also vary widely.  Thermal emission can be stable on the order of hours or days, but many outbursts or quasi-periodic pulsations (QPP) require better {than 0.1-second} time resolution, and the typical eruptive event will develop through a series of outbursts during a time interval of less than 10 minutes.  This means that although aperture synthesis could be used for the quiet Sun, it is not an option for solar radio burst science, for which instantaneous imaging is required.

The various heliospheric phenomena  are associated with radio emission covering a wide range of frequencies. The frequency range from 50 MHz to 3 GHz (or higher) suffices to monitor a wide range of solar phenomena of fundamental interest, in particular relating to flare physics, CME initiation, shock propagation, and energetic particle acceleration.  In this Chapter we present several research topics that would specifically benefit from the high resolution and sensitivity to be provided by the SKA, and which are expected to bring us new results of transformative nature.

\section{Magnetic reconnection diagnostics}\label{sec:reconn}

Solar flares produce high-energy radiation, non-thermal energetic particles and (sometimes) clouds of magnetised plasma known as Coronal Mass Ejections (CMEs) \citep[e.g.][]{2008LRSP....5....1B,fletcher11}. Each of these are major events of space weather, and can have significant effects on technological systems on Earth and on satellites in the terrestrial environment \citep{schwenn06}. They are increasingly identified as a major societal risk, with high economic impact. Radio observations with the SKA could play an important role in elucidating the underlying processes in flares, building towards predictive capacity, as well as in space weather monitoring. {The latter issue requires the SKA operation planning team to dedicate several time slots to heliospheric observations daily.} Furthermore, understanding solar flares is a challenging problem in fundamental plasma physics, with implications for other astrophysical transients including stellar flares, which have recently been shown to be  common across many types of  stars \citep{maehara12}.

Flares almost certainly represent the release of stored magnetic energy through the process of magnetic reconnection \citep{priest02, shibata11}. However, there are many unsolved problems concerning the physical processes involved. The reconnection in a flare may be triggered by a filament eruption, as in the \lq\lq standard\rq\rq\ flare model - or by another process such as newly-emerging flux or an instability. One crucial task is thus to determine the magnetic field configuration. However, the coronal magnetic field cannot be directly measured and is usually inferred by e.g. a force-free extrapolation from measured fields at the photosphere \citep{derosa09}. There are a number of difficulties with this approach even for steady fields, such as the non-force-free nature of the lower atmospheric layers, and it certainly does not  apply to rapidly-changing fields during a flare.  Indeed, current quantitative estimates of the changes in magnetic energy during a flare are inconsistent with the overall energy budget \citep{sun12}, probably due to these uncertainties in field extrapolations. Analysis of  gyrosynchrotron emission during flares can be used to constrain the coronal magnetic field \citep{fleishman09} and so studies combining the high spatial and spectral resolution of the SKA will provide far more detailed information about the magnetic field.

The magnetic reconnection site in flares cannot be directly observed, and indeed, the scale lengths of the dissipation region are likely to be extremely small, probably of the order of tens of metres. However, in order to understand the energy-release processes, it is essential to determine the location and spatial extent of this site. In the standard model this is a single monolithic current sheet, but there is increasingly strong evidence that the reconnection site is fragmented, either into a chain of magnetic islands or plasmoids, or into  even more complex three-dimensional (3D) structures \citep{cargill12,gordovskyy13}, and reconnection and turbulence interact on many scales in flares \citep{browning13}.
Non-thermal particles produced during flares are a crucial tool for indirectly observing the reconnection site \citep{zharkova11}, and are discussed further in Section \ref{sec:accel}. Their diagnostic capabilities will benefit greatly from the improved spatial resolution offered by the SKA.

\section{Quasi-periodic pulsations in flares}\label{sec:qpp}

\begin{figure}[h]
\centering
\includegraphics[height=0.3\hsize]{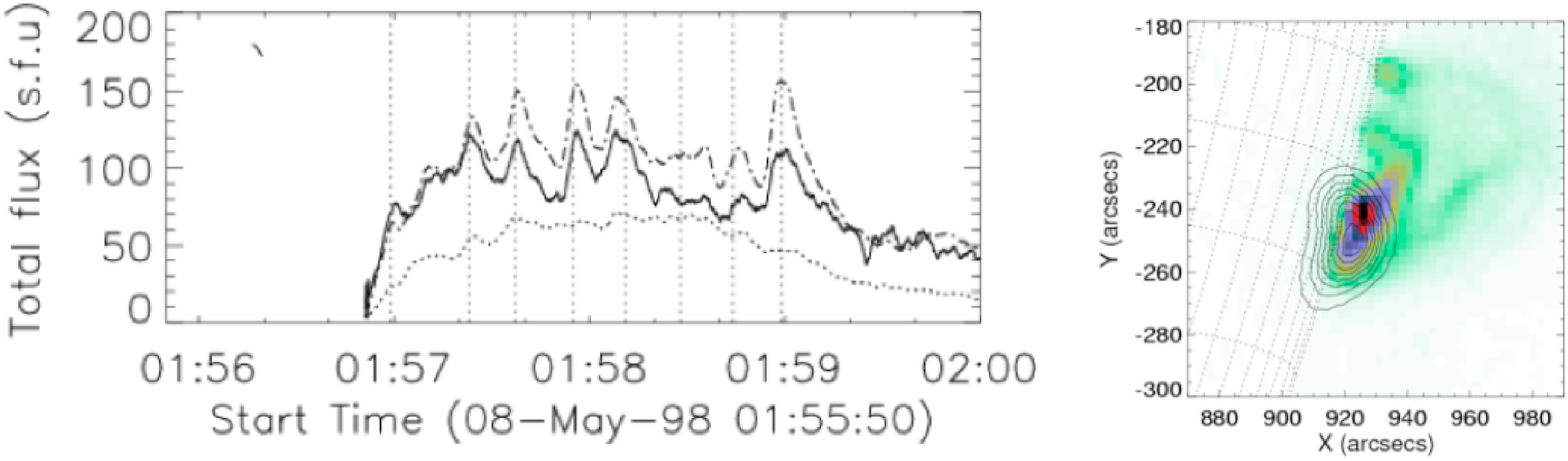}
\caption{Quasi-periodic pulsations in the solar flare of 8th May 1998. \textit{Left:} Solar flux time profiles of microwave emission at 9.4 GHz (dot-dashed line), 17 GHz (solid line) and 3.75 GHz (dotted line). \textit{Right:} Soft X-ray image from Yohkoh. Contour: microwave emission source for the same flare, observed by NoRH at 17 GHz intensity  \cite[from][]{2008A&A...487.1147I}.}
\label{qpp}
\end{figure}
The radio, white light, X-ray and gamma-ray light curves of solar and stellar flares are often found to have pronounced quasi-oscillatory patterns called quasi-periodic pulsations  \citep[QPP; see Fig.~\ref{qpp}, and ][for a comprehensive review]{2009SSRv..149..119N}. The detected periodicities range from a fraction of a second to tens of minutes and the modulation depth of the main flaring signal varies significantly between events, from a few percent to a hundred percent.

The strong correlations typically seen between the  microwave and X-ray emission indicate that the phenomenon is associated with  non-thermal electrons (Section \ref{sec:accel}), but the physical mechanisms for QPP in flares are still uncertain. Theoretical studies show that the detected periodicities may be caused by MHD oscillations of the flaring active regions or nearby plasma structures. This interpretation is supported by the recent discovery of ubiquitous MHD wave activity in the solar corona
\citep[see][for a recent review]{2012RSPTA.370.3193D}, as space and time resolved MHD waves and oscillations in coronal plasma structures in the EUV band have periods in the same range as long-period QPP. However, the shorter-period QPP ($<1$~min) are not time-resolved in the EUV band.

Coronal MHD waves are intensively used for diagnostics of the plasma in MHD seismology \citep[e.g.][]{2012PhyU...55A...4S}. The confident identification of MHD oscillations in flaring QPP will lead to the transformative change in the application of this method, as the superior time resolution intrinsic to the radio band allows the detection of phenomena on the time scale of the transverse Alfv\'en transit time, of the order of one second. Such observations
require the combination of high sensitivity with high spatial and time resolution, and hence are a natural task for the SKA. Moreover, the broad spectral coverage would allow simultaneous resolution of the processes associated with the non-thermal electrons propagating from the flare site downwards into dense plasma, seen in the high-frequency band ($>1$~GHz), and upwards, into rarified plasma, seen in the low-frequency band, again a transformative improvement for coronal seismology.

The knowledge gained from the study of QPP in solar flares can be applied to the interpretation of QPP in stellar flares, providing us with an important tool for MHD seismology of stellar coronae. Short-period
QPP have been confidently detected in stellar flares in the radio band \citep[e.g.][]{2004AstL...30..319Z}; and even occasionally in the white light and X-ray bands, despite the lack of the necessary resolution and sensitivity.  This issue demonstrates clearly the synergetic potential of this research - the SKA sensitivity will allow radical improvements to the detection of QPP in stellar flares and hence the statistical significance of the results.

\section{Particle acceleration and transport in solar flares}\label{sec:accel}




Solar flares are known
to efficiently accelerate electrons in large numbers \citep{2011SSRv..159..107H}, but the detailed physics of the process is not known. At high frequencies (above a few GHz), flare radio emission is often dominated by gyrosynchrotron radiation from such energetic ($\simeq 100$ keV) electrons in flaring loops \citep[e.g.][for a recent review]{2011SSRv..159..225W}. As noted in Section \ref{sec:reconn}, this emission can be used to constrain the coronal magnetic field, while non-thermal particles are a crucial tool to diagnose the flare reconnection site. New observations \citep[e.g.][]{2011ApJ...731L..19F} suggest that radio emission can even be used as a unique tool to diagnose
the region where energetic electrons are accelerated when traditional X-ray techniques are insensitive. In addition, optically
thin gyrosynchrotron emission provides a powerful tool to infer the poorly understood properties of the energetic electrons. Spatially and spectrally resolved observations are key to this understanding. The two fixed frequencies (17 and 34~GHz)
currently offered by NoRH, are not sufficient to disentangle features of the acceleration from the effects of electron transport. The SKA will allow the simultaneous imaging of various regions of the flaring atmosphere, and therefore significantly improve our understanding of these processes, and by extension magnetic reconnection.

Radio emission is also seen with rather short (sub-second) duration and very narrow bandwidth \citep[][as a review]{2008LRSP....5....1B}. Although in a few such cases, the spatial relation between coronal X-ray sources and such coherent radio emissions (called decimetric spikes) has been investigated,  the origin and the driver of such bursts is largely unknown. Metric spikes are also sometimes associated with the acceleration of electrons \citep[e.g.][]{2001A&A...371..333P} which lead to Type III radio bursts (see below). High sensitivity observations can substantially improve our knowledge of radio signatures, often associated with microscopic processes in the solar atmosphere \citep[e.g.][]{2004A&A...417..325K}.



Non-thermal electrons are also observed to escape the solar atmosphere and travel into interplanetary space, producing radio bursts of several kinds \citep[e.g.][as a review]{2008A&ARv..16....1P}. Radio observations at frequencies of around 50-300~MHz therefore provide a unique tool to probe the high solar corona and are often the only means by which to observe such escaping particles as well as CMEs and shocks (discussed in the next section) in this region \citep[e.g.][]{1998ARA&A..36..131B}.

Although we have a basic physical picture of electron transport from the Sun to the Earth, there are still many open questions concerning
energetic electron acceleration, storage, and release in the corona, and transport in interplanetary
space \citep[e.g.][and references therein]{2009ApJ...695L.140K}. So called Type III solar radio bursts can be produced by the propagating electrons at the local plasma frequency which ranges from hundreds of MHz in the high corona to kHz near and beyond the Earth \citep[e.g.][]{2014RAA....14..773S}. A Type III burst observed at steadily decreasing frequency thus implies a population of fast electrons injected onto an open magnetic field line in the corona and travelling outwards.
SKA observations of radio signatures at decimetric/metric wavelengths thus provide essential tools for studying not only
energetic particles, but also
the magnetic field geometry near flares and magnetic connections from the flare site to the interplanetary medium.

\section{Shocks and particle acceleration in the solar atmosphere}\label{sec:cmes}

CMEs are spectacular eruptions of plasma and magnetic field from the surface of the Sun into the heliosphere, which can travel at speeds of up to 2,500~km~s$^{-1}$ and have masses $\sim$10$^{15}$~g. CMEs often produce shocks in the solar atmosphere and thereby accelerate electrons and other particles into interplanetary space. The MHz and GHz radio emission from these accelerated electrons can be used to diagnose the acceleration mechanism, which may include magnetic reconnection and coronal shock waves \citep[e.g.][]{2004ASSL..314.....G}, although the precise details of these mechanisms remain unclear.
The SKA will enable us to address these questions using its extremely high sensitivity, spectral coverage, and imaging capabilities. These results will not only give us a new insight into the fundamental physics of CMEs and CME shocks, but will enable us to improve the forecasting of adverse space weather at Earth.


Recently, \cite{2013NatPh...9..811C} studied shocks and particle acceleration associated with the eruption of a CME
using EUV, radio and white-light imaging. The CME-induced shock was coincident with a coronal (\lq\lq EIT\rq\rq) wave and an intense metric radio burst (known as a Type II solar radio burst) generated by intermittent acceleration of electrons to kinetic energies of 2-46~keV (0.1-0.4c). Their results indicate that CME-driven quasi-perpendicular shocks are capable of producing quasi-periodic acceleration of electrons, an effect consistent with a turbulent or rippled plasma shock surface.

\begin{figure}[h]
\includegraphics[height=\hsize,angle =-90]{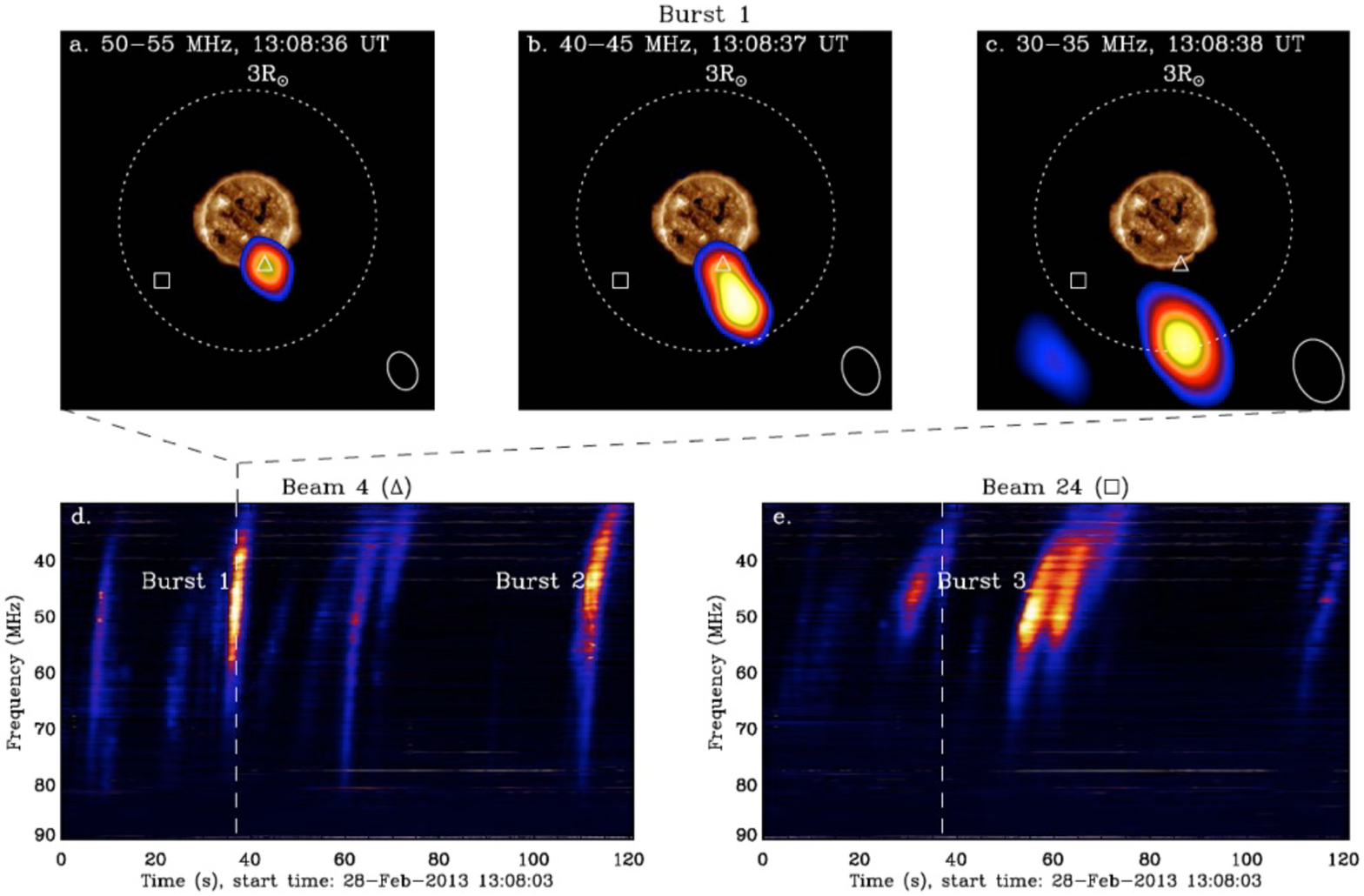}
\caption{LOFAR tied-array images of a Type III radio burst at (a) 50-55 MHz (b) 40-45 MHz and (c) 30-35 MHz, separated by 1~s. The inset is an SDO/AIA (19.3~nm) image of the EUV Sun. The dynamic spectra corresponding to two different beams, Beam 4 and Beam 24, are shown in panels d and e. The full evolution of the radio sources in the dynamic spectra (Bursts 1, 2 and 3) can be viewed in \cite{2014A&A...568A..67M}}
\label{fig2}
\end{figure}

The Nan\c{c}ay Radioheliograph used in that study provides imaging at a few discrete frequencies in the 10's to 100's of MHz, while the Birr Callisto radio-spectrometer \citep[e.g.][]{2012SoPh..280..591Z} offers good spectral coverage, but the combined superior spatial and spectral resolution offered by the SKA would radically advance these topics.
For example, the high spatial, spectral and temporal resolution offered by LOFAR \citep{2013A&A...556A...2V} was exploited by
\cite{2014A&A...568A..67M}
to study solar Type III radio bursts and their association with CMEs (see Figure~\ref{fig2}).
The non-radial high altitude Type III bursts were associated with the expanding flank of a CME which may have compressed neighbouring streamer plasma producing larger electron densities at high altitudes, while the non-radial burst trajectories can be explained by the deflection of radial magnetic fields as the CME expanded in the low corona.
New discoveries, such as the short-duration, fast-drift metric bursts recently found with the MWA \citep[e.g.][and references therein]{2011ApJ...728L..27O} also demonstrate the potential of the SKA.

Newly-available theory and simulation capabilities offer great promise to explain the detailed properties of Type II and III bursts.
For instance, impressive agreement exists between the observed and predicted dynamic spectra of several type II bursts below about 15\,MHz \citep{2014JGRA..119...69S}.  These combine 3D MHD simulation of a particular CME moving through the event-specific corona (using the BATS-R-US code with data-driven models) with detailed emission physics, and show multiple radio sources distributed across the 3D shock front.  Type III simulations have also been developed which can predict the basic properties of emission including its fine structures in frequency,
associated bremsstrahlung X-rays in the chromosphere \citep{2013A&A...550A..51H}, and intensification when the electron beam crosses a shock \citep{2012ApJ...753..124L}.  Fully 3D theories for type III bursts, as well as enhanced theories for decimetric events \citep{2011sswh.book..267C} are expected to be available before SKA1 is operational.

The SKA therefore offers the ability to address  some of the many fundamental and important issues, both observational and theoretical, for solar radio bursts \citep[see also the related discussion for the MWA;][]{2013PASA...30...31B}, by allowing high time, spatial and spectral resolution imaging. Correlations with data from other wavelengths, such as EUV and X-ray images of solar flares, magnetic reconnection, and CMEs from multiple spacecraft (e.g. SDO, STEREO, and RHESSI) and also detailed comparison with new theory and simulations will allow new and unanticipated discoveries, as well as illuminating such issues as the detailed source structure and sizes of Type II and III bursts, their evolution, polarisation characteristics, and their relation with other processes during flares.

\section{Interplanetary Scintillation}\label{sec:ips}

Interplanetary scintillation (IPS) is caused by small-scale ($\sim$80\,km to 500\,km) density variations in the solar wind crossing the line of sight (LOS) from distant, compact astronomical radio sources (or occasionally, spacecraft beacons) to the radio receiver and scattering their radio signal \citep[e.g.][and references therein]{1964Natur.203.1214H,2010SoPh..265...49B}.  Such measurements led to fundamental discoveries of solar wind structuring and the first heliospheric remote-sensing results, only confirmed much later by spacecraft.  IPS of a compact radio source gives fast (typically $\sim$0.1\,Hz to $\sim$20\,Hz) fluctuations of radio intensity/amplitude and phase, containing information about both solar wind and radio source.
When simultaneous observations of IPS are undertaken at a range of observing frequencies, say $f$ to $5f$, (i.e. say 60\,MHz to 300\,MHz -- a capability of SKA1-LOW), such data would be extremely useful in examining the scale size of density irregularities, shape of the density and cross-frequency spectra.
A mathematical description of IPS is available in
e.g. \citet{1967ApJ...147..433S,2006GeoRL..3311106F,Bisi06MOD} and references therein.

\begin{figure}[h]
\begin{center}
\includegraphics[width=\hsize]{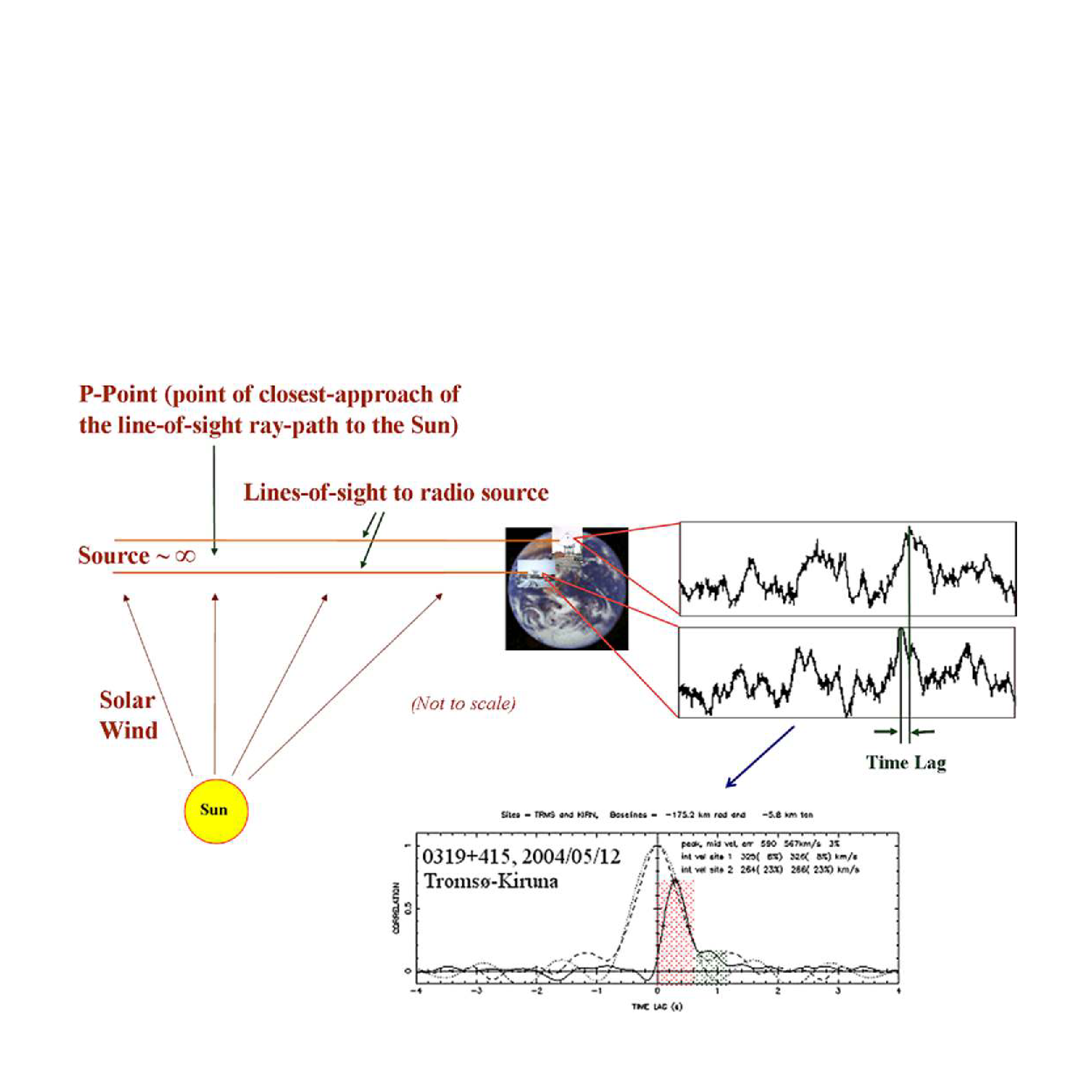}
\end{center}
\caption{The basic principles of multi-site (in this case two European Incoherent SCATter, EISCAT, radar antennas, located in Troms\o\ and Kiruna) IPS through the simultaneous observation of a single radio source as described in the text.  The signal's variation in amplitude as recorded is directly related to turbulence and density variation in the solar-wind outflow crossing the LOS.  A cross-correlation analysis of the two simultaneous signals received yields what is known as a cross-correlation function (CCF) which can be used as a first estimate of the velocity or velocities crossing the LOS.  (From \citet{2010SoPh..261..149B}).}
\label{Multi_Site_IPS_Overview}
\end{figure}

Simultaneous observations at a range of observing frequencies, for example 60\,MHz to 300\,MHz with SKA1-LOW, would be extremely useful in examining the scale size of density irregularities, shape of the density and cross-frequency spectra, and studying the transition between weak and strong scattering of the interplanetary (IP) medium.
These effects are also important in relation to other radio observations, such as Type II and III bursts  (Section \ref{sec:cmes}) as scattering can strongly affect their source sizes and directionality, and smear out any fine structure within the source. Density irregularities can also change the frequency fine structuring of Type III emission \citep{2012SoPh..279..173L}.
The smallest scale size of density irregularities (in the case of IPS, the cut-off scale or inner-scale size \citep[e.g.][]{1978JGR....83.1413C}) present in the solar wind will determine the typical scale size of radio emission. Further from the Sun this scale will be within reach with SKA1-LOW, but closer in, it may require the MID or SURVEY ranges.
More robust methods for obtaining solar wind parameters are generally based on the cross-correlation of two simultaneous observations of the same radio source \citep[e.g.][and references therein]{1996Ap&SS.243...87C,1996JATP...58..507B,2010SoPh..261..149B}.  Figure \ref{Multi_Site_IPS_Overview} provides a simplified overview of a multi-site IPS experiment.  Typically both receivers operate at the same frequency, but
\citet{Bisi06MOD} and \citet{2006GeoRL..3311106F} have shown that multi-frequency cross-correlations are also possible and can produce much interesting science (as described earlier in this section).  The SKA can potentially open up such possibilities in investigating the turbulence scale and spectrum of the solar wind.

In addition, multiple observations of IPS over several weeks or months will allow for the 3D tomographic reconstruction of the detailed structure of solar-wind outflows in the heliosphere, using, for example, the
University of California, San Diego (UCSD) algorithms \citep[e.g.][and references therein]{2013SoPh..285..151J}.
SKA1-LOW will allow a sufficient number of observations per day to reconstruct the temporal evolution of the outflows in unprecedented detail, and the application of SKA2 in the future
will allow us to access new physics on the turbulence-scale of the solar wind (scales that are currently too small to be sampled fast enough with current and upcoming in-situ instrumentation), as well as improved and higher-resolution 3D reconstructions of the inner heliosphere to allow us to better visualise and reconstruct a complete picture of the solar wind as it propagates from the Sun to the Earth's orbit and beyond.

\section{Heliospheric Faraday Rotation}\label{sec:faraday}

Space weather forecasting is of considerable interest socioeconomically, particularly for highly geo-effective events like CMEs. The impact of these depends strongly on
the presence of a southwardly-directed magnetic field, either in the CME itself or behind the CME driven shock (in the CME sheath region).  Hence, information on the magnetic field topology within a CME, especially the strength of the $B_z$ component, as it is usually referred to, is necessarily required to determine its geo-effectiveness.  At present, our earliest information of {\bf B} within a CME comes from in-situ measurements from space-based observatories located in orbit about the Sun-Earth L$_1$ point.   However, these offer at best tens of minutes early warning, and tend to be disrupted by the CME (or indeed the shocked plasma ahead of the fast CMEs).


Faraday rotation (FR) of background linearly-polarised sources allows remote sensing of the magnitude and orientation of magnetic field along the observing LOS.
The technique was pioneered using the telemetry signals from Pioneer 6 and Helios spacecraft \citep[e.g][and references therein]{1970SoPh...14..440S,1985SoPh...98..341B}, and remains a highly productive approach \citep[e.g.][and references therein]{2013SoPh..285...83J}.  Astronomical radio sources may also be used to detect coronal FR \citep[e.g.][and references therein]{2000ApJ...539..480M,2009IAUS..257..529S}.

Applied to CMEs, a large number of background sources should be observed
along the projected path of the CME in the sky plane
before, during, and after the passage of a CME, while tracking the CME as it ploughs its way out.
Each measurement provides information only on the
LOS component of {\bf B},  integrated along the entire LOS.
However, the observations of a large number ($>100$) of lines of sight through the CME as it evolves and travels through the heliosphere provide a large number of independent constraints.

Moreover, the currently-favoured flux-rope based models of CME magnetic fields require only $\sim$10 degrees of freedom in even the most general models. Therefore, self-consistent modelling of the observed FR, especially when exploiting the continuities along time and space axes, in the time-evolving CME model will yield a very-tightly constrained model for the CME flux rope.  In fact, using the time series of FR observed along a single LOS (provided by satellite beacons) \citep[][and references therein]{2008GeoRL..35.2103J,2010SoPh..265...49B} have successfully constrained the flux-rope orientation, position, size, velocity, rate of change of rope radius and pitch angle.  Multiple LOS measurements will allow much more detailed characterisation.

Indeed, in anticipation of instruments such as LOFAR and the MWA, \citet{2007ApJ...665.1439L} and \citet{2010SoPh..265...31J} showed that simultaneous observations along multiple lines of sight can indeed be used to uniquely determine the {\bf B} field in the CME flux rope.  If the turn-around time on this analysis can be reduced to a few hours, this technique can provide an early warning one-to-two days in advance, as opposed to the present warning time of a few hours at best.

FR is quadratically dependent on wavelength, and so is easier
to discern at lower frequencies.  However, at lower frequencies
the fractional polarisation of most extra-galactic sources drops significantly, the Galactic background becomes steadily brighter \citep[e.g.][]{2008AJ....136..641R} and the FR due to the terrestrial ionosphere also increases.  Additional complications arise because the lines of sight passing close to the active Sun are required, and the solar radio emission is strongly and rapidly varying and shows strong spectral features. Nonetheless, recent instruments such as the
MWA and LOFAR do intend to explore this possibility.

We expect there to be a sufficiently-dense grid of sources radiating linearly-polarised light to serve as a background grid to against which to observe the CME plasma.
 WSRT observations in the range of 340\,MHz to 370\,MHz found extra-galactic sources with typical linearly-polarised fluxes of 20\,mJy and readily measurable rotation measures (RMs) with an angular density of about one suitable polarised source every $\sim$4 square degrees \citep[e.g.][and references therein]{2003A&A...404..233H}.  These observations also found the Galactic synchrotron background emission to be significantly linearly polarised.  Recent work with the new-generation instruments has confirmed the presence of this practically ubiquitous diffuse polarised background emission at even lower frequencies \citep[e.g.][and references therein]{2014arXiv1407.2093J}. Simultaneous (or near simultaneous)
observations of IPS alongside those of heliospheric FR will also provide
additional context information on the structure(s) within the heliosphere to enable a better
understanding of what exactly is causing the FR or the changes in FR recorded say, during
the passage of a CME.

The optimal observing frequencies lie in the upper part of the SKA1-LOW range.
In practise, triggered observing will be required for these hard-to-predict events,
most likely using a space-based observatory, which can provide information of the launch time, direction, and speed of a CME.  Heliospheric imagers, such as those aboard the
twin STEREO spacecraft, can provide additional information determining the patch of sky to be monitored.
The observations ideally require good time resolution,
and angular resolution of order tens of arc seconds to an arc minute
in the spectral line mode, with a spectral resolution of a few 10s of kHz.  To maximise the benefit from the use of technique of RM synthesis \citep[e.g.][]{2005A&A...441.1217B}, these observations should cover the widest bandwidth possible. Contamination from the ionospheric FR signal is a problem over much of the region of interest, and will require calibrating down to a few percent, but \citet{2013A&A...552A..58S} have already shown good progress in this respect.  The SKA1-LOW projected sensitivity and dynamic range will offer exciting new observations using this technique, both for forecasting, and for the improved understanding of CMEs in general.



\section{Summary}

The impact and usefulness of SKA1 for coronal observations will depend on the imaging capabilities available and their adaptation for solar observations.  The first criterion to consider is the frequency coverage and the type of instrument.  SKA1-LOW will cover part of the range of LOFAR, with a denser
coverage for the short baselines, which is an important factor for solar imaging since LOFAR is somewhat sparse in the tens of meters to a few hundred meter baselines necessary to define the shape of the solar disk at meter wavelengths.  SKA1-LOW, if observing during daytime, will need to include in its operating mode a way to deal with the rather high fluxes originating from solar outbursts so solar imaging using SKA-LOW is more likely than with SKA1-SUR and SKA1-MID which we will discuss later.  Unfortunately SKA1-LOW will be placed far from the instrument which offers the best complement in terms of frequency range, SKA1-MID. This means that although SKA1-MID and SKA1-LOW together will cover the frequency range from 50\,MHz to 3,050\,MHz a given solar radio outburst which is a short feature on the scale of minutes or seconds will either be observed by one or the other but not both.  Simultaneity may not always be possible due to the Earth's longitudinal separation of the two antenna types. However, solar observations with SKA1-LOW will well overlap in time with the NoRH, and the upcoming high-frequency spectral radio heliographs of the new generation, CSRH and SSRT.

SKA1-SUR can still provide some complementary measures, but the fact that it provides lower resolution with a much sparser field of view than SKA1-MID and in particular the fact that it will not be upgraded for SKA2 mean that SKA1-MID plus LOFAR would jointly outperform SKA1-SUR plus SKA1-LOW.  We thus expect SKA1-LOW to essentially overlap the solar science case for LOFAR, likely with better image performance for the Sun and with NoRH, CSRH and SSRT as complementary instruments at higher frequencies.  This in particular opens up an interesting possibility for simultaneous tracing of the emission of
non-thermal particles going up and down from a flare site.

Since for solar purposes the case for SKA1-SUR is reasonably similar to the case for SKA1-MID and since SKA1-MID will the only one to be significantly upgraded for SKA2 we concentrate the further discussion on SKA-MID.  The frequency range range for SKA1-MID, from 350\,MHz to 3,050\,MHz, means we could study the range from 450\,MHz to 1\,GHz where there has been a noticeable lack of imaging on the Sun. That alone would be a major contribution for solar physics from SKA1. We note a couple of issues on the possibility of imaging with SKA1-MID as defined for SKA1.

First there is the fact that the dish size (15 metres) and the single beam means that at the highest frequencies the antenna field of view is smaller than the angular size of the Sun (the synthetic field of view will be even smaller).  Below 1~GHz the size of the field of view should be adequate if particular attention is taken to ensure that the short baseline coverage is enough for the Sun to be effectively imaged out to about two solar radii from the Sun's centre. This field of view aspect will be less of an issue if the SKA1-MID dishes are equipped with phased array feeds and probably will not be an issue for SKA2 if dense aperture arrays are to be implemented.

The greatest problem in using SKA1-MID to observe the solar corona relates to the very large fluxes involved. Solar radiastronomers define their flux in terms of solar flux units (SFU) such that 1\,SFU\,=\,10,000\,Jy. Solar fluxes in the decimeter wavelength range are often close to 100 SFU, and in large events such as the Halloween 2003 events, the solar emission in the decimeter to meter wavelength can exceed 10$^5$ SFU \citep{2005ApJ...631L..97P}. The dense aperture array systems to be installed during SKA2, if performing operations during daytime, will need to handle such fluxes in their normal operating modes, so even major solar outbursts can readily be included in the science topics for SKA2. The dishes that will be used for SKA1 do not need to handle solar like fluxes even during daytime operations since the observing schedule can always be defined in order to maintain the Sun outside of the antenna field of view. If pointed at the Sun there will be a major departure from the sub-Jansky modes typical of non-solar observations. On the other hand, observations of IPS will be possible using SKA1-MID, and FR could also be attempted within respectable elongation angles from the Sun, and we feel that this, together with imaging observations with SKA1-LOW, atop of the IPS and FR prospects with SKA1-LOW, will be the most important contributions from SKA1 for Heliophysics including solar and space weather studies.

Despite some challenges and potential cost implications  involved with coronal imaging using SKA1-MID, we note that such high fluxes must be handled during any  daytime operation by the dense aperture arrays during SKA2. Also, the multi-beam characteristics of the dense aperture arrays will then facilitate allocation of time for solar operation.  IPS observations with SKA1-MID are not affected by these issues, and these, alongside imaging observations with SKA1-LOW, will certainly make important contributions from SKA1 to radio heliophysics and space weather studies.



We conclude that the observational parameters of both SKA1 and SKA2 will provide a major breakthrough in solar and heliospheric radiophysics, bringing new discoveries and qualitatively advancing solar and heliospheric physics, fundamental plasma astrophysics, and space weather.

\bibliographystyle{apj}
\bibliography{vmn_bib_files,Heliosphere_Refs}

\begin{thebibliography}{}
\expandafter\ifx\csname natexlab\endcsname\relax\def\natexlab#1{#1}\fi

\bibitem[{{Bastian} {et~al.}(1998){Bastian}, {Benz}, \&
  {Gary}}]{1998ARA&A..36..131B}
{Bastian}, T.~S., {Benz}, A.~O., \& {Gary}, D.~E. 1998, \araa, 36, 131

\bibitem[{{Benz}(2008)}]{2008LRSP....5....1B}
{Benz}, A.~O. 2008, Living Reviews in Solar Physics, 5, 1

\bibitem[{{Bird} {et~al.}(1985){Bird}, {Volland}, {Howard}, {Koomen},
  {Michels}, {Sheeley}, {Amstrong}, {Seidel}, {Stelzried}, \&
  {Woo}}]{1985SoPh...98..341B}
{Bird}, M.~K., {Volland}, H., {Howard}, R.~A., {et~al.} 1985, \solphys, 98, 341

\bibitem[{Bisi(2006)}]{Bisi06MOD}
Bisi, M.~M. 2006, {Ph.D. Thesis
  (http://cadair.aber.ac.uk/dspace/handle/2160/4064)}, {The University of
  Wales, Aberystwyth}

\bibitem[{{Bisi} {et~al.}(2010{\natexlab{a}}){Bisi}, {Fallows}, {Breen}, \&
  {O'Neill}}]{2010SoPh..261..149B}
{Bisi}, M.~M., {Fallows}, R.~A., {Breen}, A.~R., \& {O'Neill}, I.~J.
  2010{\natexlab{a}}, \solphys, 261, 149

\bibitem[{{Bisi} {et~al.}(2010{\natexlab{b}}){Bisi}, {Breen}, {Jackson},
  {Fallows}, {Walsh}, {Miki{\'c}}, {Riley}, {Owen}, {Gonzalez-Esparza},
  {Aguilar-Rodriguez}, {Morgan}, {Jensen}, {Wood}, {Owens}, {Tokumaru},
  {Manoharan}, {Chashei}, {Giunta}, {Linker}, {Shishov}, {Tyul'Bashev},
  {Agalya}, {Glubokova}, {Hamilton}, {Fujiki}, {Hick}, {Clover}, \&
  {Pint{\'e}r}}]{2010SoPh..265...49B}
{Bisi}, M.~M., {Breen}, A.~R., {Jackson}, B.~V., {et~al.} 2010{\natexlab{b}},
  \solphys, 265, 49

\bibitem[{{Bowman} {et~al.}(2013){Bowman}, {Cairns}, {Kaplan}, {Murphy},
  {Oberoi}, {Staveley-Smith}, {Arcus}, {Barnes}, {Bernardi}, {Briggs}, {Brown},
  {Bunton}, {Burgasser}, {Cappallo}, {Chatterjee}, {Corey}, {Coster},
  {Deshpande}, {deSouza}, {Emrich}, {Erickson}, {Goeke}, {Gaensler},
  {Greenhill}, {Harvey-Smith}, {Hazelton}, {Herne}, {Hewitt},
  {Johnston-Hollitt}, {Kasper}, {Kincaid}, {Koenig}, {Kratzenberg}, {Lonsdale},
  {Lynch}, {Matthews}, {McWhirter}, {Mitchell}, {Morales}, {Morgan}, {Ord},
  {Pathikulangara}, {Prabu}, {Remillard}, {Robishaw}, {Rogers}, {Roshi},
  {Salah}, {Sault}, {Shankar}, {Srivani}, {Stevens}, {Subrahmanyan}, {Tingay},
  {Wayth}, {Waterson}, {Webster}, {Whitney}, {Williams}, {Williams}, \&
  {Wyithe}}]{2013PASA...30...31B}
{Bowman}, J.~D., {Cairns}, I., {Kaplan}, D.~L., {et~al.} 2013, \pasa, 30, 31

\bibitem[{{Braun}(1997)}]{1997LNP...483..207B}
{Braun}, R. 1997, in Lecture Notes in Physics, Berlin Springer Verlag, Vol.
  483, Coronal Physics from Radio and Space Observations, ed. G.~{Trottet}, 207

\bibitem[{{Breen} {et~al.}(1996){Breen}, {Coles}, {Grall}, {Lovhaug},
  {Markkanen}, {Misawa}, \& {Williams}}]{1996JATP...58..507B}
{Breen}, A.~R., {Coles}, W.~A., {Grall}, R., {et~al.} 1996, Journal of
  Atmospheric and Terrestrial Physics, 58, 507

\bibitem[{{Brentjens} \& {de Bruyn}(2005)}]{2005A&A...441.1217B}
{Brentjens}, M.~A., \& {de Bruyn}, A.~G. 2005, \aap, 441, 1217

\bibitem[{{Browning} \& {Lazarian}(2013)}]{browning13}
{Browning}, P., \& {Lazarian}, A. 2013, \ssr, 178, 325

\bibitem[{{Cairns}(2011)}]{2011sswh.book..267C}
{Cairns}, I.~H. 2011, {Coherent Radio Emissions Associated with Star System
  Shocks}, ed. M.~P. {Miralles} \& J.~{S{\'a}nchez Almeida}, 267

\bibitem[{{Cargill} {et~al.}(2012){Cargill}, {Vlahos}, {Baumann}, {Drake}, \&
  {Nordlund}}]{cargill12}
{Cargill}, P.~J., {Vlahos}, L., {Baumann}, G., {Drake}, J.~F., \& {Nordlund},
  {\AA}. 2012, \ssr, 173, 223

\bibitem[{{Carley} {et~al.}(2013){Carley}, {Long}, {Byrne}, {Zucca},
  {Bloomfield}, {McCauley}, \& {Gallagher}}]{2013NatPh...9..811C}
{Carley}, E.~P., {Long}, D.~M., {Byrne}, J.~P., {et~al.} 2013, Nature Physics,
  9, 811

\bibitem[{{Coles}(1996)}]{1996Ap&SS.243...87C}
{Coles}, W.~A. 1996, \apss, 243, 87

\bibitem[{{Coles} \& {Harmon}(1978)}]{1978JGR....83.1413C}
{Coles}, W.~A., \& {Harmon}, J.~K. 1978, \jgr, 83, 1413

\bibitem[{{De Moortel} \& {Nakariakov}(2012)}]{2012RSPTA.370.3193D}
{De Moortel}, I., \& {Nakariakov}, V.~M. 2012, Royal Society of London
  Philosophical Transactions Series A, 370, 3193

\bibitem[{{De Rosa} {et~al.}(2009){De Rosa}, {Schrijver}, {Barnes}, {Leka},
  {Lites}, {Aschwanden}, {Amari}, {Canou}, {McTiernan}, {R{\'e}gnier},
  {Thalmann}, {Valori}, {Wheatland}, {Wiegelmann}, {Cheung}, {Conlon},
  {Fuhrmann}, {Inhester}, \& {Tadesse}}]{derosa09}
{De Rosa}, M.~L., {Schrijver}, C.~J., {Barnes}, G., {et~al.} 2009, \apj, 696,
  1780

\bibitem[{{Fallows} {et~al.}(2006){Fallows}, {Breen}, {Bisi}, {Jones}, \&
  {Wannberg}}]{2006GeoRL..3311106F}
{Fallows}, R.~A., {Breen}, A.~R., {Bisi}, M.~M., {Jones}, R.~A., \& {Wannberg},
  G. 2006, \grl, 33, 11106

\bibitem[{{Fleishman} {et~al.}(2011){Fleishman}, {Kontar}, {Nita}, \&
  {Gary}}]{2011ApJ...731L..19F}
{Fleishman}, G.~D., {Kontar}, E.~P., {Nita}, G.~M., \& {Gary}, D.~E. 2011,
  \apjl, 731, L19

\bibitem[{{Fleishman} {et~al.}(2009){Fleishman}, {Nita}, \&
  {Gary}}]{fleishman09}
{Fleishman}, G.~D., {Nita}, G.~M., \& {Gary}, D.~E. 2009, \apjl, 698, L183

\bibitem[{{Fletcher} {et~al.}(2011){Fletcher}, {Dennis}, {Hudson}, {Krucker},
  {Phillips}, {Veronig}, {Battaglia}, {Bone}, {Caspi}, {Chen}, {Gallagher},
  {Grigis}, {Ji}, {Liu}, {Milligan}, \& {Temmer}}]{fletcher11}
{Fletcher}, L., {Dennis}, B.~R., {Hudson}, H.~S., {et~al.} 2011, \ssr, 159, 19

\bibitem[{{Gary} \& {Keller}(2004)}]{2004ASSL..314.....G}
{Gary}, D.~E., \& {Keller}, C.~U., eds. 2004, Astrophysics and Space Science
  Library, Vol. 314, {Solar and Space Weather Radiophysics - Current Status and
  Future Developments}

\bibitem[{{Gary} {et~al.}(2012){Gary}, {Nita}, \& {Sane}}]{2012AAS...22020430G}
{Gary}, D.~E., {Nita}, G.~M., \& {Sane}, N. 2012, in American Astronomical
  Society Meeting Abstracts, Vol. 220, American Astronomical Society Meeting
  Abstracts \#220, 204.30

\bibitem[{{Gordovskyy} {et~al.}(2013){Gordovskyy}, {Browning}, {Kontar}, \&
  {Bian}}]{gordovskyy13}
{Gordovskyy}, M., {Browning}, P.~K., {Kontar}, E.~P., \& {Bian}, N.~H. 2013,
  \solphys, 284, 489

\bibitem[{{Hannah} {et~al.}(2013){Hannah}, {Kontar}, \&
  {Reid}}]{2013A&A...550A..51H}
{Hannah}, I.~G., {Kontar}, E.~P., \& {Reid}, H.~A.~S. 2013, \aap, 550, A51

\bibitem[{{Haverkorn} {et~al.}(2003){Haverkorn}, {Katgert}, \& {de
  Bruyn}}]{2003A&A...404..233H}
{Haverkorn}, M., {Katgert}, P., \& {de Bruyn}, A.~G. 2003, \aap, 404, 233

\bibitem[{{Hewish} {et~al.}(1964){Hewish}, {Scott}, \&
  {Wills}}]{1964Natur.203.1214H}
{Hewish}, A., {Scott}, P.~F., \& {Wills}, D. 1964, \nat, 203, 1214

\bibitem[{{Holman} {et~al.}(2011){Holman}, {Aschwanden}, {Aurass}, {Battaglia},
  {Grigis}, {Kontar}, {Liu}, {Saint-Hilaire}, \&
  {Zharkova}}]{2011SSRv..159..107H}
{Holman}, G.~D., {Aschwanden}, M.~J., {Aurass}, H., {et~al.} 2011, \ssr, 159,
  107

\bibitem[{{Inglis} {et~al.}(2008){Inglis}, {Nakariakov}, \&
  {Melnikov}}]{2008A&A...487.1147I}
{Inglis}, A.~R., {Nakariakov}, V.~M., \& {Melnikov}, V.~F. 2008, \aap, 487,
  1147

\bibitem[{{Jackson} {et~al.}(2013){Jackson}, {Clover}, {Hick}, {Buffington},
  {Bisi}, \& {Tokumaru}}]{2013SoPh..285..151J}
{Jackson}, B.~V., {Clover}, J.~M., {Hick}, P.~P., {et~al.} 2013, \solphys, 285,
  151

\bibitem[{{Jelic} {et~al.}(2014){Jelic}, {de Bruyn}, {Mevius}, {Abdalla},
  {Asad}, {Bernardi}, {Brentjens}, {Bus}, {Chapman}, {Ciardi}, {Daiboo},
  {Fernandez}, {Ghosh}, {Harker}, {Jensen}, {Kazemi}, {Koopmans},
  {Labropoulos}, {Martinez-Rubi}, {Mellema}, {Offringa}, {Pandey}, {Patil},
  {Thomas}, {Vedantham}, {Veligatla}, {Yatawatta}, {Zaroubi}, {Alexov},
  {Anderson}, {Avruch}, {Beck}, {Bell}, {Bentum}, {Best}, {Bonafede},
  {Bregman}, {Breitling}, {Broderick}, {Brouw}, {Bruggen}, {Butcher}, {Conway},
  {de Gasperin}, {de Geus}, {Deller}, {Dettmar}, {Duscha}, {Eisloffel},
  {Engels}, {Falcke}, {Fallows}, {Fender}, {Ferrari}, {Frieswijk}, {Garrett},
  {Griessmeier}, {Gunst}, {Hamaker}, {Hassall}, {Haverkorn}, {Heald},
  {Hessels}, {Hoeft}, {Horandel}, {Horneffer}, {van der Horst}, {Iacobelli},
  {Juette}, {Karastergiou}, {Kondratiev}, {Kramer}, {Kuniyoshi}, {Kuper}, {van
  Leeuwen}, {Maat}, {Mann}, {McKay-Bukowski}, {McKean}, {Munk}, {Nelles},
  {Norden}, {Paas}, {Pandey-Pommier}, {Pietka}, {Pizzo}, {Polatidis}, {Reich},
  {Rottgering}, {Rowlinson}, {Scaife}, {Schwarz}, {Serylak}, {Smirnov},
  {Steinmetz}, {Stewart}, {Tagger}, {Tang}, {Tasse}, {ter Veen}, {Thoudam},
  {Toribio}, {Vermeulen}, {Vocks}, {van Weeren}, {Wijers}, {Wijnholds},
  {Wucknitz}, \& {Zarka}}]{2014arXiv1407.2093J}
{Jelic}, V., {de Bruyn}, A.~G., {Mevius}, M., {et~al.} 2014, ArXiv e-prints,
  arXiv:1407.2093

\bibitem[{{Jensen} {et~al.}(2013){Jensen}, {Bisi}, {Breen}, {Heiles}, {Minter},
  \& {Vilas}}]{2013SoPh..285...83J}
{Jensen}, E.~A., {Bisi}, M.~M., {Breen}, A.~R., {et~al.} 2013, \solphys, 285,
  83

\bibitem[{{Jensen} {et~al.}(2010){Jensen}, {Hick}, {Bisi}, {Jackson}, {Clover},
  \& {Mulligan}}]{2010SoPh..265...31J}
{Jensen}, E.~A., {Hick}, P.~P., {Bisi}, M.~M., {et~al.} 2010, \solphys, 265, 31

\bibitem[{{Jensen} \& {Russell}(2008)}]{2008GeoRL..35.2103J}
{Jensen}, E.~A., \& {Russell}, C.~T. 2008, \grl, 35, 2103

\bibitem[{{Karlick{\'y}}(2004)}]{2004A&A...417..325K}
{Karlick{\'y}}, M. 2004, \aap, 417, 325

\bibitem[{{Kontar} \& {Reid}(2009)}]{2009ApJ...695L.140K}
{Kontar}, E.~P., \& {Reid}, H.~A.~S. 2009, \apjl, 695, L140

\bibitem[{{Lesovoi} {et~al.}(2014){Lesovoi}, {Altyntsev}, {Ivanov}, \&
  {Gubin}}]{2014RAA....14..864L}
{Lesovoi}, S.~V., {Altyntsev}, A.~T., {Ivanov}, E.~F., \& {Gubin}, A.~V. 2014,
  Research in Astronomy and Astrophysics, 14, 864

\bibitem[{{Li} \& {Cairns}(2012)}]{2012ApJ...753..124L}
{Li}, B., \& {Cairns}, I.~H. 2012, \apj, 753, 124

\bibitem[{{Li} {et~al.}(2012){Li}, {Cairns}, \&
  {Robinson}}]{2012SoPh..279..173L}
{Li}, B., {Cairns}, I.~H., \& {Robinson}, P.~A. 2012, \solphys, 279, 173

\bibitem[{{Liu} {et~al.}(2007){Liu}, {Manchester}, {Kasper}, {Richardson}, \&
  {Belcher}}]{2007ApJ...665.1439L}
{Liu}, Y., {Manchester}, IV, W.~B., {Kasper}, J.~C., {Richardson}, J.~D., \&
  {Belcher}, J.~W. 2007, \apj, 665, 1439

\bibitem[{{Maehara} {et~al.}(2012){Maehara}, {Shibayama}, {Notsu}, {Notsu},
  {Nagao}, {Kusaba}, {Honda}, {Nogami}, \& {Shibata}}]{maehara12}
{Maehara}, H., {Shibayama}, T., {Notsu}, S., {et~al.} 2012, \nat, 485, 478

\bibitem[{{Mancuso} \& {Spangler}(2000)}]{2000ApJ...539..480M}
{Mancuso}, S., \& {Spangler}, S.~R. 2000, \apj, 539, 480

\bibitem[{{Mann} {et~al.}(2011){Mann}, {Vocks}, \&
  {Breitling}}]{2011pre7.conf..507M}
{Mann}, G., {Vocks}, C., \& {Breitling}, F. 2011, Planetary, Solar and
  Heliospheric Radio Emissions (PRE VII), 507

\bibitem[{{Morosan} {et~al.}(2014){Morosan}, {Gallagher}, {Zucca}, {Fallows},
  {Carley}, {Mann}, {Bisi}, {Kerdraon}, {Konovalenko}, {MacKinnon}, {Rucker},
  {Thid{\'e}}, {Magdaleni{\'c}}, {Vocks}, {Reid}, {Anderson}, {Asgekar},
  {Avruch}, {Bentum}, {Bernardi}, {Best}, {Bonafede}, {Bregman}, {Breitling},
  {Broderick}, {Br{\"u}ggen}, {Butcher}, {Ciardi}, {Conway}, {de Gasperin}, {de
  Geus}, {Deller}, {Duscha}, {Eisl{\"o}ffel}, {Engels}, {Falcke}, {Ferrari},
  {Frieswijk}, \& {Garrett}}]{2014A&A...568A..67M}
{Morosan}, D.~E., {Gallagher}, P.~T., {Zucca}, P., {et~al.} 2014, \aap, 568,
  A67

\bibitem[{{Nakajima} {et~al.}(1994){Nakajima}, {Nishio}, {Enome}, {Shibasaki},
  {Takano}, {Hanaoka}, {Torii}, {Sekiguchi}, {Bushimata}, {Kawashima},
  {Shinohara}, {Irimajiri}, {Koshiishi}, {Kosugi}, {Shiomi}, {Sawa}, \&
  {Kai}}]{1994IEEEP..82..705N}
{Nakajima}, H., {Nishio}, M., {Enome}, S., {et~al.} 1994, IEEE Proceedings, 82,
  705

\bibitem[{{Nakariakov} \& {Melnikov}(2009)}]{2009SSRv..149..119N}
{Nakariakov}, V.~M., \& {Melnikov}, V.~F. 2009, \ssr, 149, 119

\bibitem[{{Oberoi} {et~al.}(2011){Oberoi}, {Matthews}, {Cairns}, {Emrich},
  {Lobzin}, {Lonsdale}, {Morgan}, {Prabu}, {Vedantham}, {Wayth}, {Williams},
  {Williams}, {White}, {Allen}, {Arcus}, {Barnes}, {Benkevitch}, {Bernardi},
  {Bowman}, {Briggs}, {Bunton}, {Burns}, {Cappallo}, {Clark}, {Corey},
  {Dawson}, {DeBoer}, {De Gans}, {deSouza}, {Derome}, {Edgar}, {Elton},
  {Goeke}, {Gopalakrishna}, {Greenhill}, {Hazelton}, {Herne}, {Hewitt},
  {Kamini}, {Kaplan}, {Kasper}, {Kennedy}, {Kincaid}, {Kocz}, {Koeing},
  {Kowald}, {Lynch}, {Madhavi}, {McWhirter}, {Mitchell}, {Morales}, {Ng},
  {Ord}, {Pathikulangara}, {Rogers}, {Roshi}, {Salah}, {Sault}, {Schinckel},
  {Udaya Shankar}, {Srivani}, {Stevens}, {Subrahmanyan}, {Thakkar}, {Tingay},
  {Tuthill}, {Vaccarella}, {Waterson}, {Webster}, \&
  {Whitney}}]{2011ApJ...728L..27O}
{Oberoi}, D., {Matthews}, L.~D., {Cairns}, I.~H., {et~al.} 2011, \apjl, 728,
  L27

\bibitem[{{Paesold} {et~al.}(2001){Paesold}, {Benz}, {Klein}, \&
  {Vilmer}}]{2001A&A...371..333P}
{Paesold}, G., {Benz}, A.~O., {Klein}, K.-L., \& {Vilmer}, N. 2001, \aap, 371,
  333

\bibitem[{{Pick} {et~al.}(2005){Pick}, {Malherbe}, {Kerdraon}, \&
  {Maia}}]{2005ApJ...631L..97P}
{Pick}, M., {Malherbe}, J.-M., {Kerdraon}, A., \& {Maia}, D.~J.~F. 2005, \apjl,
  631, L97

\bibitem[{{Pick} \& {Vilmer}(2008)}]{2008A&ARv..16....1P}
{Pick}, M., \& {Vilmer}, N. 2008, \aapr, 16, 1

\bibitem[{{Priest} \& {Forbes}(2002)}]{priest02}
{Priest}, E.~R., \& {Forbes}, T.~G. 2002, \aapr, 10, doi:10.1007/s001590100013

\bibitem[{{Rogers} \& {Bowman}(2008)}]{2008AJ....136..641R}
{Rogers}, A.~E.~E., \& {Bowman}, J.~D. 2008, \aj, 136, 641

\bibitem[{{Salpeter}(1967)}]{1967ApJ...147..433S}
{Salpeter}, E.~E. 1967, \apj, 147, 433

\bibitem[{{Schmidt} \& {Cairns}(2014)}]{2014JGRA..119...69S}
{Schmidt}, J.~M., \& {Cairns}, I.~H. 2014, Journal of Geophysical Research
  (Space Physics), 119, 69

\bibitem[{{Schwenn}(2006)}]{schwenn06}
{Schwenn}, R. 2006, Living Reviews in Solar Physics, 3, 2

\bibitem[{{Shibata} \& {Magara}(2011)}]{shibata11}
{Shibata}, K., \& {Magara}, T. 2011, Living Reviews in Solar Physics, 8, 6

\bibitem[{{Sinclair Reid} \& {Ratcliffe}(2014)}]{2014RAA....14..773S}
{Sinclair Reid}, H.~A., \& {Ratcliffe}, H. 2014, Research in Astronomy and
  Astrophysics, 14, 773

\bibitem[{{Sotomayor-Beltran} {et~al.}(2013){Sotomayor-Beltran}, {Sobey},
  {Hessels}, {de Bruyn}, {Noutsos}, {Alexov}, {Anderson}, {Asgekar}, {Avruch},
  {Beck}, {Bell}, {Bell}, {Bentum}, {Bernardi}, {Best}, {Birzan}, {Bonafede},
  {Breitling}, {Broderick}, {Brouw}, {Br{\"u}ggen}, {Ciardi}, {de Gasperin},
  {Dettmar}, {van Duin}, {Duscha}, {Eisl{\"o}ffel}, {Falcke}, {Fallows},
  {Fender}, {Ferrari}, {Frieswijk}, {Garrett}, {Grie{\ss}meier}, {Grit},
  {Gunst}, {Hassall}, {Heald}, {Hoeft}, {Horneffer}, {Iacobelli}, {Juette},
  {Karastergiou}, {Keane}, {Kohler}, {Kramer}, {Kondratiev}, {Koopmans},
  {Kuniyoshi}, {Kuper}, {van Leeuwen}, {Maat}, {Macario}, {Markoff}, {McKean},
  {Mulcahy}, {Munk}, {Orru}, {Paas}, {Pandey-Pommier}, {Pilia}, {Pizzo},
  {Polatidis}, {Reich}, {R{\"o}ttgering}, {Serylak}, {Sluman}, {Stappers},
  {Tagger}, {Tang}, {Tasse}, {ter Veen}, {Vermeulen}, {van Weeren}, {Wijers},
  {Wijnholds}, {Wise}, {Wucknitz}, {Yatawatta}, \&
  {Zarka}}]{2013A&A...552A..58S}
{Sotomayor-Beltran}, C., {Sobey}, C., {Hessels}, J.~W.~T., {et~al.} 2013, \aap,
  552, A58

\bibitem[{{Spangler} \& {Whiting}(2009)}]{2009IAUS..257..529S}
{Spangler}, S.~R., \& {Whiting}, C.~A. 2009, in IAU Symposium, Vol. 257, IAU
  Symposium, ed. N.~{Gopalswamy} \& D.~F. {Webb}, 529--541

\bibitem[{{Stelzried} {et~al.}(1970){Stelzried}, {Levy}, {Sato}, {Rusch},
  {Ohlson}, {Schatten}, \& {Wilcox}}]{1970SoPh...14..440S}
{Stelzried}, C.~T., {Levy}, G.~S., {Sato}, T., {et~al.} 1970, \solphys, 14, 440

\bibitem[{{Stepanov} {et~al.}(2012){Stepanov}, {Zaitsev}, \&
  {Nakariakov}}]{2012PhyU...55A...4S}
{Stepanov}, A.~V., {Zaitsev}, V.~V., \& {Nakariakov}, V.~M. 2012, Physics
  Uspekhi, 55, A4

\bibitem[{{Sun} {et~al.}(2012){Sun}, {Hoeksema}, {Liu}, {Wiegelmann},
  {Hayashi}, {Chen}, \& {Thalmann}}]{sun12}
{Sun}, X., {Hoeksema}, J.~T., {Liu}, Y., {et~al.} 2012, \apj, 748, 77

\bibitem[{{van Haarlem} {et~al.}(2013){van Haarlem}, {Wise}, {Gunst}, {Heald},
  {McKean}, {Hessels}, {de Bruyn}, {Nijboer}, {Swinbank}, {Fallows},
  {Brentjens}, {Nelles}, {Beck}, {Falcke}, {Fender}, {H{\"o}randel},
  {Koopmans}, {Mann}, {Miley}, {R{\"o}ttgering}, {Stappers}, {Wijers},
  {Zaroubi}, {van den Akker}, {Alexov}, {Anderson}, {Anderson}, {van Ardenne},
  {Arts}, {Asgekar}, {Avruch}, {Batejat}, {B{\"a}hren}, {Bell}, {Bell}, {van
  Bemmel}, {Bennema}, {Bentum}, {Bernardi}, {Best}, {B{\^i}rzan}, {Bonafede},
  {Boonstra}, {Braun}, {Bregman}, {Breitling}, {van de Brink}, {Broderick},
  {Broekema}, {Brouw}, {Br{\"u}ggen}, {Butcher}, {van Cappellen}, {Ciardi},
  {Coenen}, {Conway}, {Coolen}, {Corstanje}, {Damstra}, {Davies}, {Deller},
  {Dettmar}, {van Diepen}, {Dijkstra}, {Donker}, {Doorduin}, {Dromer}, {Drost},
  {van Duin}, {Eisl{\"o}ffel}, {van Enst}, {Ferrari}, {Frieswijk}, {Gankema},
  {Garrett}, {de Gasperin}, {Gerbers}, {de Geus}, {Grie{\ss}meier}, {Grit},
  {Gruppen}, {Hamaker}, {Hassall}, {Hoeft}, {Holties}, {Horneffer}, {van der
  Horst}, {van Houwelingen}, {Huijgen}, {Iacobelli}, {Intema}, {Jackson},
  {Jelic}, {de Jong}, {Juette}, {Kant}, {Karastergiou}, {Koers}, {Kollen},
  {Kondratiev}, {Kooistra}, {Koopman}, {Koster}, {Kuniyoshi}, {Kramer},
  {Kuper}, {Lambropoulos}, {Law}, {van Leeuwen}, {Lemaitre}, {Loose}, {Maat},
  {Macario}, {Markoff}, {Masters}, {McFadden}, {McKay-Bukowski}, {Meijering},
  {Meulman}, {Mevius}, {Middelberg}, {Millenaar}, {Miller-Jones}, {Mohan},
  {Mol}, {Morawietz}, {Morganti}, {Mulcahy}, {Mulder}, {Munk}, {Nieuwenhuis},
  {van Nieuwpoort}, {Noordam}, {Norden}, {Noutsos}, {Offringa}, {Olofsson},
  {Omar}, {Orr{\'u}}, {Overeem}, {Paas}, {Pandey-Pommier}, {Pandey}, {Pizzo},
  {Polatidis}, {Rafferty}, {Rawlings}, {Reich}, {de Reijer}, {Reitsma},
  {Renting}, {Riemers}, {Rol}, {Romein}, {Roosjen}, {Ruiter}, {Scaife}, {van
  der Schaaf}, {Scheers}, {Schellart}, {Schoenmakers}, {Schoonderbeek},
  {Serylak}, {Shulevski}, {Sluman}, {Smirnov}, {Sobey}, {Spreeuw}, {Steinmetz},
  {Sterks}, {Stiepel}, {Stuurwold}, {Tagger}, {Tang}, {Tasse}, {Thomas},
  {Thoudam}, {Toribio}, {van der Tol}, {Usov}, {van Veelen}, {van der Veen},
  {ter Veen}, {Verbiest}, {Vermeulen}, {Vermaas}, {Vocks}, {Vogt}, {de Vos},
  {van der Wal}, {van Weeren}, {Weggemans}, {Weltevrede}, {White}, {Wijnholds},
  {Wilhelmsson}, {Wucknitz}, {Yatawatta}, {Zarka}, {Zensus}, \& {van
  Zwieten}}]{2013A&A...556A...2V}
{van Haarlem}, M.~P., {Wise}, M.~W., {Gunst}, A.~W., {et~al.} 2013, \aap, 556,
  A2

\bibitem[{{White} {et~al.}(2011){White}, {Benz}, {Christe},
  {F{\'a}rn{\'{\i}}k}, {Kundu}, {Mann}, {Ning}, {Raulin}, {Silva-V{\'a}lio},
  {Saint-Hilaire}, {Vilmer}, \& {Warmuth}}]{2011SSRv..159..225W}
{White}, S.~M., {Benz}, A.~O., {Christe}, S., {et~al.} 2011, \ssr, 159, 225

\bibitem[{{Yan} {et~al.}(2009){Yan}, {Zhang}, {Wang}, {Liu}, {Chen}, \&
  {Ji}}]{2009EM&P..104...97Y}
{Yan}, Y., {Zhang}, J., {Wang}, W., {et~al.} 2009, Earth Moon and Planets, 104,
  97

\bibitem[{{Zaitsev} {et~al.}(2004){Zaitsev}, {Kislyakov}, {Stepanov}, {Kliem},
  \& {Furst}}]{2004AstL...30..319Z}
{Zaitsev}, V.~V., {Kislyakov}, A.~G., {Stepanov}, A.~V., {Kliem}, B., \&
  {Furst}, E. 2004, Astronomy Letters, 30, 319

\bibitem[{{Zharkova} {et~al.}(2011){Zharkova}, {Arzner}, {Benz}, {Browning},
  {Dauphin}, {Emslie}, {Fletcher}, {Kontar}, {Mann}, {Onofri}, {Petrosian},
  {Turkmani}, {Vilmer}, \& {Vlahos}}]{zharkova11}
{Zharkova}, V.~V., {Arzner}, K., {Benz}, A.~O., {et~al.} 2011, \ssr, 159, 357

\bibitem[{{Zucca} {et~al.}(2012){Zucca}, {Carley}, {McCauley}, {Gallagher},
  {Monstein}, \& {McAteer}}]{2012SoPh..280..591Z}
{Zucca}, P., {Carley}, E.~P., {McCauley}, J., {et~al.} 2012, \solphys, 280, 591

\end{thebibliography}

\end{document}